\documentclass{PoS}
\usepackage{psfrag,epsfig,graphicx,graphics,xcolor}

\newcommand{\veck}{{\bf k}}

\newcommand{\veckone}{{\bf k}_1}
\newcommand{\vecktwo}{{\bf k}_2}
\newcommand{\veckj}{{\bf k}_{J}}

\newcommand{\veckjone}{{\bf k}_{J,1}}
\newcommand{\veckjtwo}{{\bf k}_{J,2}}

\newcommand{\deins}[1]{{\rm d}#1\,}
\newcommand{\dzwei}[1]{{\rm d}^2#1\,}
\newcommand{\dk}{\dzwei{\veck}}

\newcommand{\dkone}{\dzwei{\veckone}}
\newcommand{\dktwo}{\dzwei{\vecktwo}}

\newcommand{\dsigma}{\deins{\sigma}}
\newcommand{\dsigmahat}{\deins{{\hat\sigma}_{\rm{ab}}}}
\newcommand{\dnu}{\deins{\nu}}

\newcommand{\dx}{\deins{x}}
\newcommand{\dxone}{\deins{x_1}}
\newcommand{\dxtwo}{\deins{x_2}}
\newcommand{\dyjetone}{\deins{y_{J,1}}}
\newcommand{\dyjettwo}{\deins{y_{J,2}}}
\newcommand{\dphij}{\deins{\phi_{J}}}
\newcommand{\dphijone}{\deins{\phi_{J,1}}}
\newcommand{\dphijtwo}{\deins{\phi_{J,2}}}
\newcommand{\dtwojets}{{\rm d}|\veckjone|\,{\rm d}|\veckjtwo|\,\dyjetone \dyjettwo}

\newcommand{\shat}{{\hat s}}

\newcommand{\asbar}{{\bar{\alpha}}_s}

\newcommand{\avgcosn}{\langle \cos n \varphi \rangle}
\newcommand{\avgcos}{\langle \cos \varphi \rangle}
\newcommand{\avgcostwo}{\langle \cos 2 \varphi \rangle}

\title{Mueller-Navelet jets at LHC: an observable to reveal high energy resummation effects?}

\ShortTitle{Mueller-Navelet jets at LHC: an observable to reveal high energy resummation effects?}

\author{\speaker{Bertrand Duclou\'e}\\
        LPT, Universit\'e Paris-Sud, CNRS, 91405, Orsay, France\\
        E-mail: \email{Bertrand.Ducloue@th.u-psud.fr}}

\author{Lech Szymanowski\\
        National Center for Nuclear Research (NCBJ), Warsaw, Poland\\
        E-mail: \email{Lech.Szymanowski@fuw.edu.pl}}

\author{Samuel Wallon\\
        LPT, Universit{\'e} Paris-Sud, CNRS, 91405, Orsay, France\\
        UPMC Univ. Paris 06, facult\'e de physique, 4 place Jussieu, 75252 Paris Cedex 05, France
        E-mail: \email{Samuel.Wallon@th.u-psud.fr}}

\abstract{More than 25 years ago, Mueller and Navelet proposed to study the production of two jets separated by a large interval of rapidity at hadron colliders to look for high-energy resummation effects. We here present the results of a next-to-leading logarithmic BFKL study of the azimuthal decorrelation of these jets. This includes next-to-leading corrections both to the Green's function and to the jet vertices. We compare our results with recent LHC data and results obtained in a fixed order treatment.}

\FullConference{Photon 2013,\\
		20-24 May 2013\\
		Paris, France }

\begin{document}

\section{Introduction}
In the high-energy limit of QCD, the smallness of the strong coupling $\alpha_s$ can be compensated by large logarithmic enhancements of the type $[\alpha_s\ln(s/|t|)]^n$ which can all be of the same order of magnitude and so have to be resummed. This resummation gives rise to the leading logarithmic (LL) Balitsky-Fadin-Kuraev-Lipatov (BFKL) Pomeron \cite{Fadin:1975cb,Kuraev:1976ge,Kuraev:1977fs,Balitsky:1978ic}. To study this limit, several processes have been suggested and studied, at $ep$, $ee$ and $pp$ colliders, from inclusive and semi-inclusive~\cite{Mueller:1990erAskew:1993rnNavelet:1996jxMunier:1998vkBartels:1996keBrodsky:1996sgBialas:1997eqBoonekamp:1998veKwiecinski:1999yxBrodsky:1998knBrodsky:2002ka}
to exclusive level~\cite{Ryskin:1992uiFrankfurt:1997fjEnberg:2002zyIvanov:2004vdIvanov:2000uqEnberg:2003jwPoludniowski:2003ykPire:2005icEnberg:2005eqSegond:2007fjIvanov:2005gnIvanov:2006gtCaporale:2007vs}. One of the most promising test was suggested by Mueller and Navelet, who proposed to study the production of two jets separated by a large interval of rapidity at hadron colliders \cite{Mueller:1986ey}. When using a pure leading order collinear approach, these two jets would be emitted back-to-back. On the contrary, a BFKL calculation allows some emission between the jets, which should lead to a larger cross section and less azimuthal correlation between the jets. We here present results of a full NLL analysis of this process, where the NLL corrections (corresponding to resumming also terms 
of the type $\alpha_s[\alpha_s\ln(s/|t|)]^n$) are included both for the BFKL Green's function~\cite{Fadin:1998py,Ciafaloni:1998gs} and the jet vertices~\cite{Bartels:2001ge,Bartels:2002yj,Caporale:2011cc}.

In the following we will focus on the azimuthal correlations $\avgcosn\equiv \langle\cos[n(\phi_{J,1}-\phi_{J,2}-\pi)]\rangle$ of the jets and ratios of these observables as they have been measured recently at a center of mass energy $\sqrt{s}=7$ at the LHC by the CMS collaboration~\cite{CMS-PAS-FSQ-12-002}. We will compare our results with these data and with the results obtained in a next-to-leading order (NLO) fixed order calculation.

\section{Basic formulas}

\begin{figure}[htbp]
\centering
\includegraphics[height=8cm]{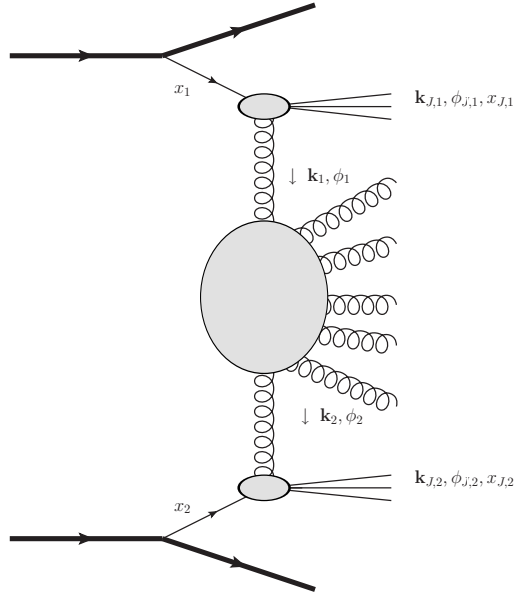}
\caption{Kinematics of the process}
\label{Fig:kinematics}
\end{figure}

The process we are studying is shown on figure~\ref{Fig:kinematics}. Two hadrons collide at a center of mass energy $\sqrt{s}$. Within collinear factorization, we can write the differential cross section as
\begin{equation}
  \frac{\dsigma}{\dtwojets} = \sum_{{\rm a},{\rm b}} \int_0^1 \dxone \int_0^1 \dxtwo f_{\rm a}(x_1) f_{\rm b}(x_2) \frac{\dsigmahat}{\dtwojets},
\end{equation}
where $\veckjone$, $\veckjtwo$ are the transverse momenta of the two jets, $y_{J,1}$ and $y_{J,2}$ their rapidities and $f_{\rm a,\, b}$ are the parton distribution functions (PDFs) of a parton a (b) in the according proton. The expression of the partonic cross section is
\begin{equation}
  \frac{\dsigmahat}{\dtwojets} = \int \dphijone\dphijtwo\int\dkone\dktwo V_{\rm a}(-\veckone,x_1)\,G(\veckone,\vecktwo,\shat)\,V_{\rm b}(\vecktwo,x_2),\label{eq:bfklpartonic}
\end{equation}
where $\phi_{J,1}$ and $\phi_{J,2}$ are the azimuthal angles of the jets, $V_{a,b}$ is the jet vertex initiated by the parton a (b) and $G$ is the BFKL Green's function which depends on $\shat=x_1 x_2 s$.
In the following, we will make use of the coefficients $\mathcal{C}_n$ defined as
\begin{equation}
  \mathcal{C}_n = (4-3\delta_{n,0}) \int \dnu C_{n,\nu}(|\veckjone|,x_{J,1})C^*_{n,\nu}(|\veckjtwo|,x_{J,2}) \left( \frac{\shat}{s_0} \right)^{\omega(n,\nu)}\,.
  \label{Cn}
\end{equation}
With this definition, $\mathcal{C}_0$ corresponds to the differential cross section:
\begin{equation}
  \mathcal{C}_0=\frac{\dsigma}{\dtwojets} \,,
%\label{def:dsigma}
\end{equation}
and the azimuthal correlations can be expressed as
\begin{equation}
  \langle\cos(n\varphi)\rangle \equiv \langle\cos\big(n(\phi_{J,1}-\phi_{J,2}-\pi)\big)\rangle = \frac{\mathcal{C}_n}{\mathcal{C}_0}\,.
\end{equation}
In eq.~(\ref{Cn}), the coefficients $C_{n,\nu}$ are defined as
\begin{equation}
   C_{n,\nu}(|\veckj|,x_{J})= \int\dphij\dk \dx f(x) V(\veck,x) E_{n,\nu}(\veck) \cos(n\phi_J)\,,
  \label{Cnnu}
\end{equation}
where $E_{n,\nu}$ are the LL BFKL eigenfunctions
\begin{equation}
  E_{n,\nu}(\veck) = \frac{1}{\pi\sqrt{2}}\left(\veck^2\right)^{i\nu-\frac{1}{2}}e^{in\phi}\,.
\label{def:eigenfunction}
\end{equation}
At LL accuracy, $\omega(n,\nu)$ is
\begin{equation}
  \omega(n,\nu) = \asbar \chi_0\left(|n|,\frac{1}{2}+i\nu\right), \quad \chi_0(n,\gamma) = 2\Psi(1)-\Psi\left(\gamma+\frac{n}{2}\right)-\Psi\left(1-\gamma+\frac{n}{2}\right)\,,
\end{equation}
with $\asbar = \alpha N_c/\pi$ and $\Psi(z) = \Gamma'(z)/\Gamma(z)\,,$
and the jet vertex reads
\begin{equation}
  V_{\rm a}(\veck,x)=V_{\rm a}^{(0)}(\veck,x) = \frac{\alpha_s}{\sqrt{2}}\frac{C_{A/F}}{\veck^2} \delta\left(1-\frac{x_J}{x}\right)|\veckj|\delta^{(2)}(\veck-\veckj)\,,
\end{equation}
($C_A$ for ${\rm a}={\rm g}$ and $C_F$ for ${\rm a}={\rm q}$), while at NLL, we have
\begin{equation}
  \omega(n,\nu) = \asbar \chi_0\left(|n|,\frac{1}{2}+i\nu\right) + \asbar^2 \left[ \chi_1\left(|n|,\frac{1}{2}+i\nu\right)-\frac{\pi b_0}{N_c}\chi_0\left(|n|,\frac{1}{2}+i\nu\right) \ln\frac{|\veckjone|\cdot|\veckjtwo|}{\mu_R^2} \right]\,,
\end{equation}
with $b_0=(33 - 2 \,N_f)/(12 \pi)$
and $V_{\rm a}(\veck,x) = V^{(0)}_{\rm a}(\veck,x) + \alpha_s V^{(1)}_{\rm a}(\veck,x)$. The expression of the NLL corrections to the Green's function resulting in $\chi_1$ can be found in eq.~(2.17) of ref.~\cite{Ducloue:2013hia}. The expressions of the NLO corrections to the jet vertices $V^{(1)}$ are quite lenghty and will not be reproduced here. They can be found in ref.~\cite{Colferai:2010wu}, as extracted from refs.~\cite{Bartels:2001ge,Bartels:2002yj} after correcting a few misprints of ref.~\cite{Bartels:2001ge}. They have been recently reobtained in ref.~\cite{Caporale:2011cc}. In the limit of small cone jets, they have been computed in ref.~\cite{Ivanov:2012ms} and applied to phenomenology in refs.~\cite{Caporale:2012ih,Caporale:2013uva}.
When using the NLO jet vertex, two partons can be emitted close to each other in the $(y,\phi)$ plane and so one should choose an appropriate jet clustering algorithm to determine if these two partons should be combined into a single jet. In the present work we will use the cone algorithm with a size parameter $R_{\rm cone}=0.5$. We have checked that using the $k_t$ or anti-$k_t$ algorithms do not change our results significantly for the observables that we will study here. In the following we will take the renormalization scale equal to the factorization scale, $\mu_R=\mu_F=\mu$. We will choose the central value $\sqrt{|\veckjone|\cdot |\veckjtwo|}$ for $\mu$ and $\sqrt{s_0}$ and vary these scales by a factor of $2$ to evaluate the scale uncertainty of our predictions. We use the MSTW 2008 PDFs \cite{Martin:2009iq} and a two-loop running coupling. We will also see how the inclusion of the collinear improvement to the Green's function, as was suggested in refs.
~\cite{Salam:1998tj,Ciafaloni:1998iv,Ciafaloni:1999yw,Ciafaloni:2003rd}
and extended for $n \neq 0$ in refs.~\cite{Vera:2007kn,Schwennsen:2007hs,Marquet:2007xx}, affects our predictions.

\section{Results for a symmetric configuration}

We will first show results in a symmetric configuration, where the lower cut on the transverse momenta of the jets is the same for both jets. We use the cuts defined below:
\begin{eqnarray}
 35\,{\rm GeV} < &|\veckjone|,|\veckjtwo|& < 60 \,{\rm GeV} \,, \nonumber\\
 0 < & y_{J,1}, \, y_{J,2} & < 4.7\,.
 \label{sym-cuts}
\end{eqnarray}
These cuts are almost the same as the ones used by the CMS collaboration in~\cite{CMS-PAS-FSQ-12-002}, except that we have to impose an upper cut on the transverse momenta of the jets to deal with the numerical integration over $|\veckjone|$ and $|\veckjtwo|$. However, we have checked that the results we will show in the following for the azimuthal correlations of the jets do not depend strongly on this cut as the cross section is quickly decreasing with increasing $|\veckjone|$ and $|\veckjtwo|$. Thus we can compare our results with LHC data.

We will consider several BFKL scenarios, starting from a pure LL approximation up to a full NLL calculation. The convention for colors that we will use for the plots showing the different treatments is the following:
\begin{equation}
\begin{tabular}{ll}
blue:    &  pure LL result (LO vertices and LL Green's function)\\
magenta: &   LO vertices and NLL Green's function \\
green:   &  LO vertices and collinear improved NLL Green's function \\
brown:   &  full NLL result (NLO vertices and NLL Green's function)\\
red:     &  NLO vertices and collinear improved NLL Green's function.
\end{tabular}
\label{def:colors}
\end{equation}

\begin{figure}[htbp]
  \def\sca{.6}
  \psfrag{central}[l][r][0.5]{\hspace{-0.9cm}pure NLL}
  \psfrag{muchange_0.5}[l][r][\sca]{\hspace{-1.8cm}\footnotesize $\mu_F \to \mu_F/2$}
  \psfrag{muchange_2.0}[l][r][\sca]{\hspace{-1.9cm} \footnotesize $\mu_F \to2 \mu_F$}
  \psfrag{s0change_0.5}[l][r][\sca]{\hspace{-1.95cm} \footnotesize $\sqrt{s_0} \to \sqrt{s_0}/2$}
  \psfrag{s0change_2.0}[l][r][\sca]{\hspace{-1.95cm} \footnotesize $\sqrt{s_0} \to 2 \sqrt{s_0}$}
  \psfrag{CMS}[l][r][0.5]{\hspace{-0.7cm} CMS data}
  \psfrag{cos}{\raisebox{.1cm}{\scalebox{0.9}{$\langle \cos \varphi\rangle$}}}
  \psfrag{Y}{\scalebox{0.9}{$Y$}}
  \begin{minipage}{0.49\textwidth}
      \includegraphics[width=7cm]{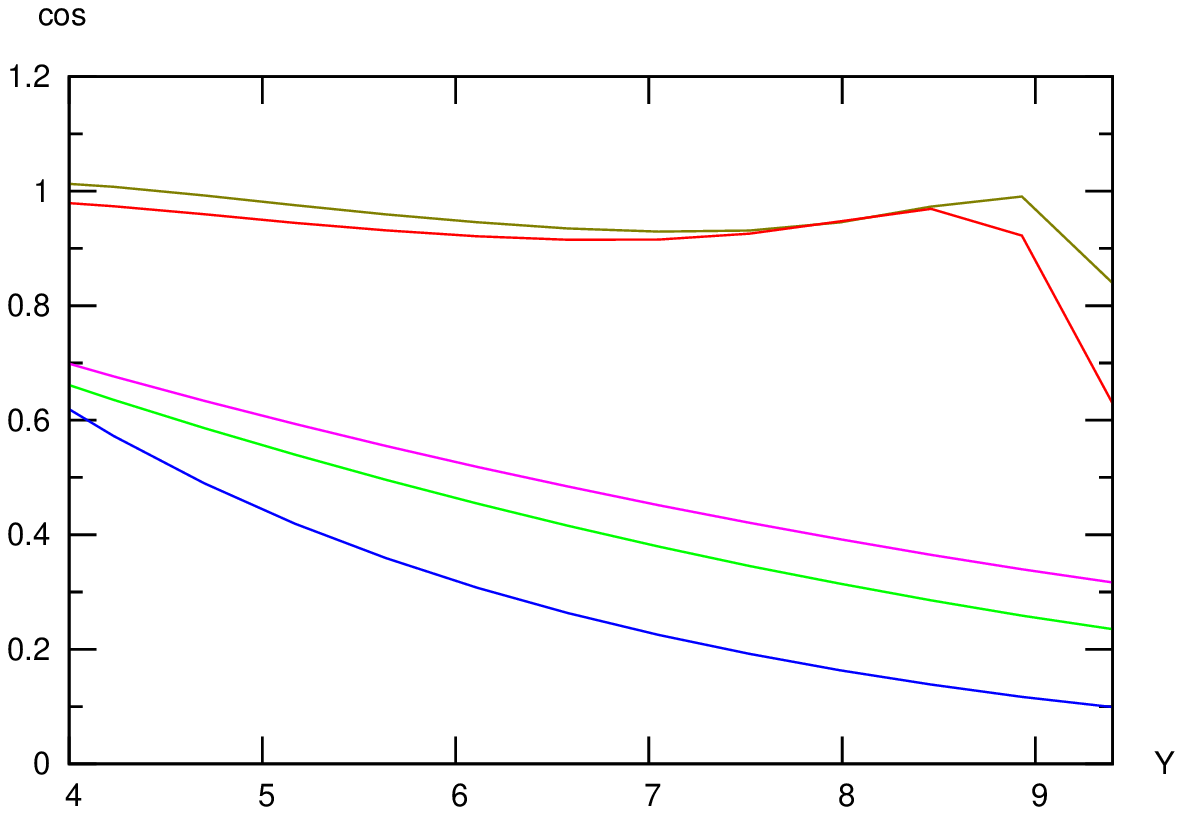}
  \end{minipage}
  \begin{minipage}{0.49\textwidth}
      \includegraphics[width=7cm]{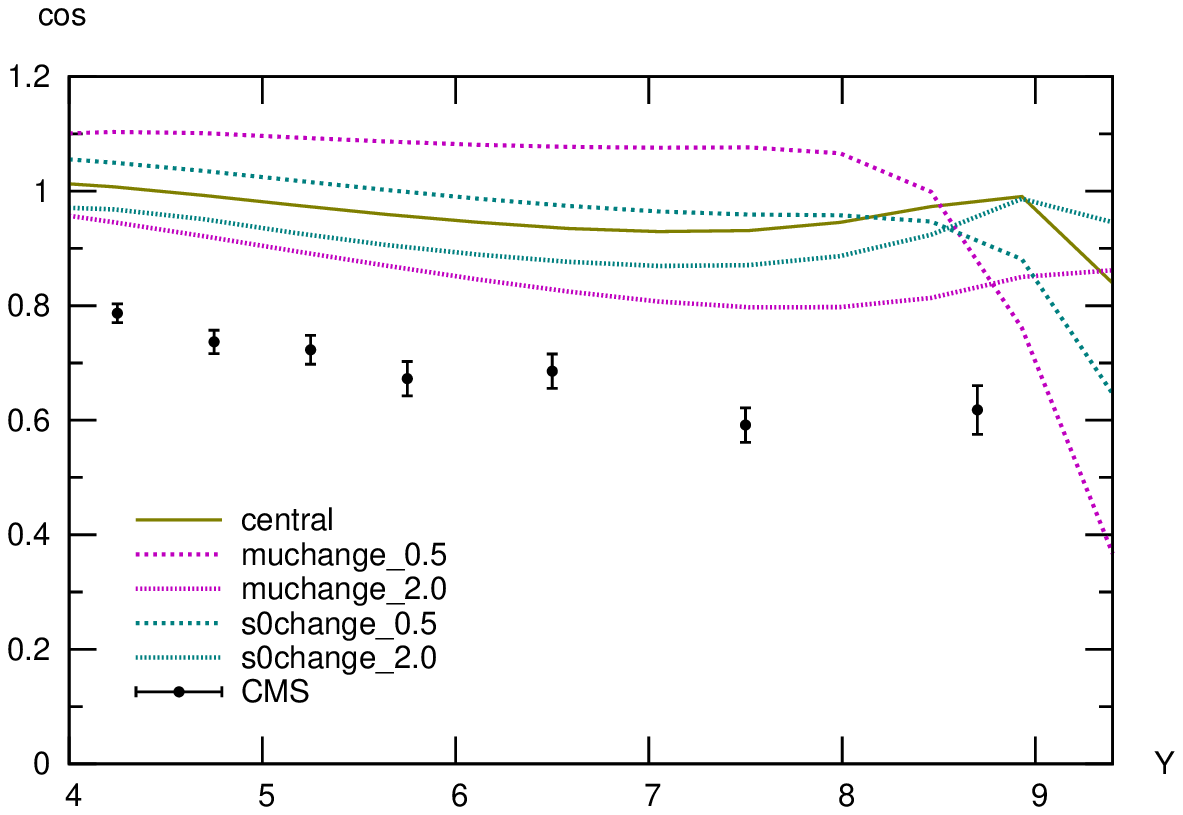}
  \end{minipage}
  \caption{Left: value of $\avgcos$ as a function of the rapidity separation $Y$, using symmetric cuts defined in (\protect\ref{sym-cuts}), for the 5 different BFKL treatments (\protect\ref{def:colors}). Right: comparison of the full NLL BFKL calculation including the scale uncertainty with CMS data (black dots with error bars).}
  \label{Fig:cos_sym}
\end{figure}

We first consider the azimuthal correlation $\avgcos$. On figure~\ref{Fig:cos_sym} (L) we show the variation of $\avgcos$ with respect to the rapidity separation between the jets $Y \equiv |y_{J,1} - y_{J,2}|$ for the 5 BFKL treatments~\ref{def:colors}. We recall that a value of $1$ corresponds to jets always emitted back-to-back, while a value of $0$ means that there is no correlation of the jets. We can see that the pure LL calculation leads to a very strong decorrelation between the two jets. The inclusion of NLL corrections to the Green's function leads to a small increase of the correlation. The NLO correction to the jet vertices have a very large impact and lead to a very strong correlation with a value of $\avgcos$ very close to $1$ and a much flatter behavior with respect to $Y$. The effect of the collinear improvement of the Green's function is sizable when using the LO vertices, and leads to a slightly smaller correlation. But when convoluted with the NLO vertices, the collinear improved NLL Green's
function gives results very similar to the one based on the pure NLL Green's function. On figure~\ref{Fig:cos_sym} (R) we show the effect of varying the scales $\mu$ and $s_0$ by a factor of 2 on our full NLL results and show a comparison with CMS data (black dots with error bars). The NLL results with the 'natural' scale choice $\mu=s_0=\sqrt{|\veckjone|\cdot |\veckjtwo|}$ predicts a too large correlation when compared to the data, but we can see that the uncertainty coming from the scale choice is quite large for this observable.

\begin{figure}[htbp]
  \def\sca{.6}
  \psfrag{central}[l][r][0.5]{\hspace{-2cm}pure NLL}
  \psfrag{muchange_0.5}[l][r][\sca]{\hspace{-1.7cm}\footnotesize $\mu_F \to \mu_F/2$}
  \psfrag{muchange_2.0}[l][r][\sca]{\hspace{-1.8cm} \footnotesize $\mu_F \to2 \mu_F$}
  \psfrag{s0change_0.5}[l][r][\sca]{\hspace{-1.95cm} \footnotesize $\sqrt{s_0} \to \sqrt{s_0}/2$}
  \psfrag{s0change_2.0}[l][r][\sca]{\hspace{-1.95cm} \footnotesize $\sqrt{s_0} \to 2 \sqrt{s_0}$}
  \psfrag{CMS}[l][r][0.5]{\hspace{-2.1cm} CMS data}
  \psfrag{cos}{\raisebox{.1cm}{\scalebox{0.9}{$\langle \cos 2 \varphi\rangle$}}}
  \psfrag{Y}{\scalebox{0.9}{$Y$}}
  \begin{minipage}{0.49\textwidth}
      \includegraphics[width=7cm]{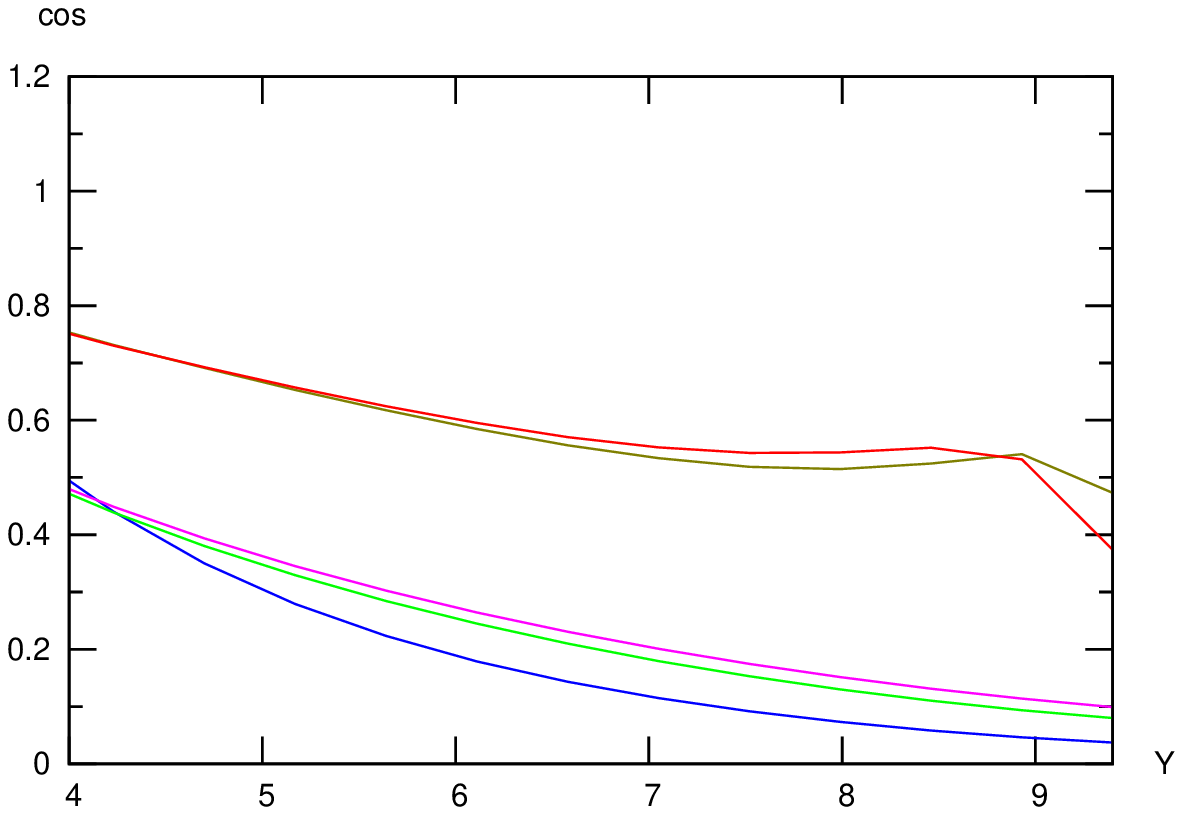}
  \end{minipage}
  \begin{minipage}{0.49\textwidth}
      \includegraphics[width=7cm]{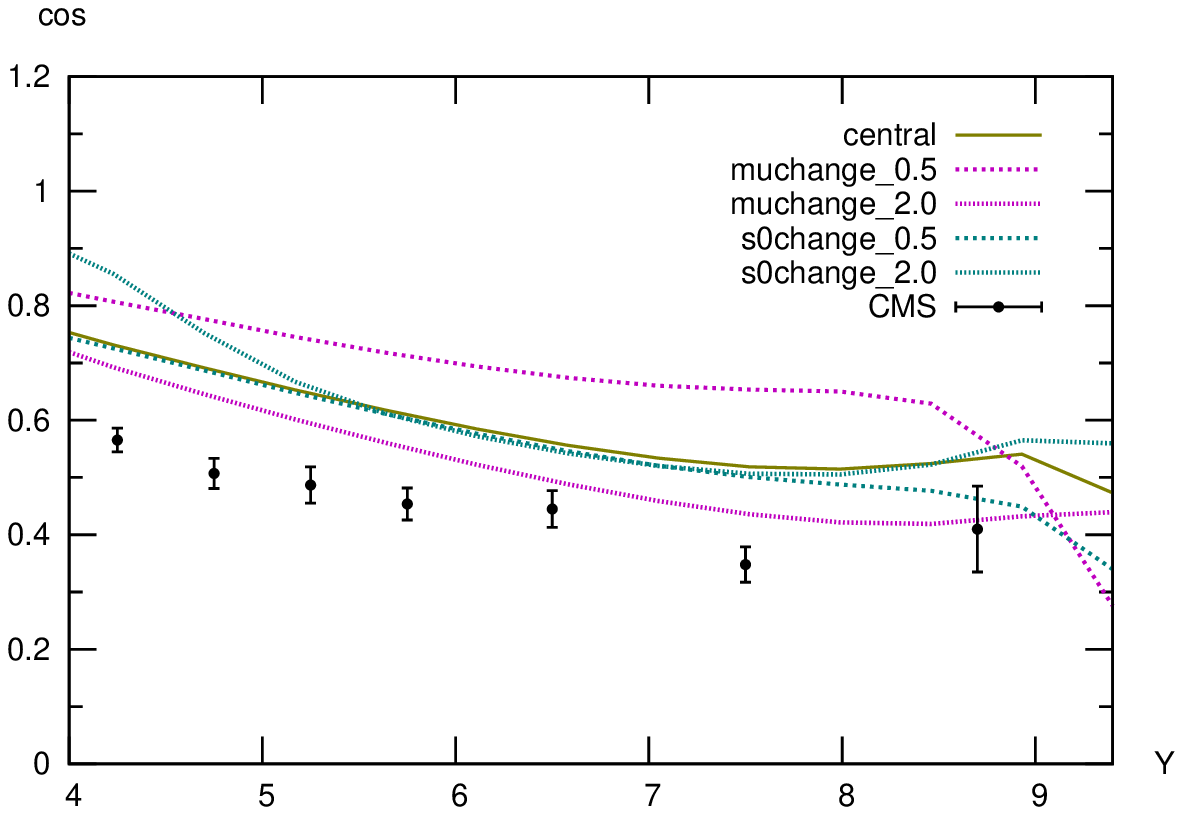}
  \end{minipage}
  \caption{Left: value of $\avgcostwo$ as a function of the rapidity separation $Y$, using symmetric cuts defined in (\protect\ref{sym-cuts}), for the 5 different BFKL treatments (\protect\ref{def:colors}). Right: comparison of the full NLL BFKL calculation including the scale uncertainty with CMS data (black dots with error bars).}
  \label{Fig:cos2_sym}
\end{figure}

In figure~\ref{Fig:cos2_sym} we perform the same analysis for the observable $\avgcostwo$. Here again the importance of NLO corrections to the jet vertices is much larger than the NLL corrections to the Green's function, and the dependency on the scales choice is quite large. We can see that when using the choice $\mu=2\sqrt{|\veckjone|\cdot |\veckjtwo|}$ the NLL BFKL calculation is not very far from the data.

\begin{figure}[htbp]
  \def\sca{.6}
  \psfrag{central}[l][r][0.5]{\hspace{-0.9cm}pure NLL}
  \psfrag{muchange_0.5}[l][r][\sca]{\hspace{-1.8cm}\footnotesize $\mu_F \to \mu_F/2$}
  \psfrag{muchange_2.0}[l][r][\sca]{\hspace{-1.9cm} \footnotesize $\mu_F \to2 \mu_F$}
  \psfrag{s0change_0.5}[l][r][\sca]{\hspace{-1.95cm} \footnotesize $\sqrt{s_0} \to \sqrt{s_0}/2$}
  \psfrag{s0change_2.0}[l][r][\sca]{\hspace{-1.95cm} \footnotesize $\sqrt{s_0} \to 2 \sqrt{s_0}$}
  \psfrag{CMS}[l][r][0.5]{\hspace{-0.7cm} CMS data}
  \psfrag{cos}{\raisebox{.1cm}{\scalebox{0.9}{$\langle \cos 2 \varphi\rangle / \langle \cos \varphi\rangle$}}}
  \psfrag{Y}{\scalebox{0.9}{$Y$}}
  \begin{minipage}{0.49\textwidth}
      \includegraphics[width=7cm]{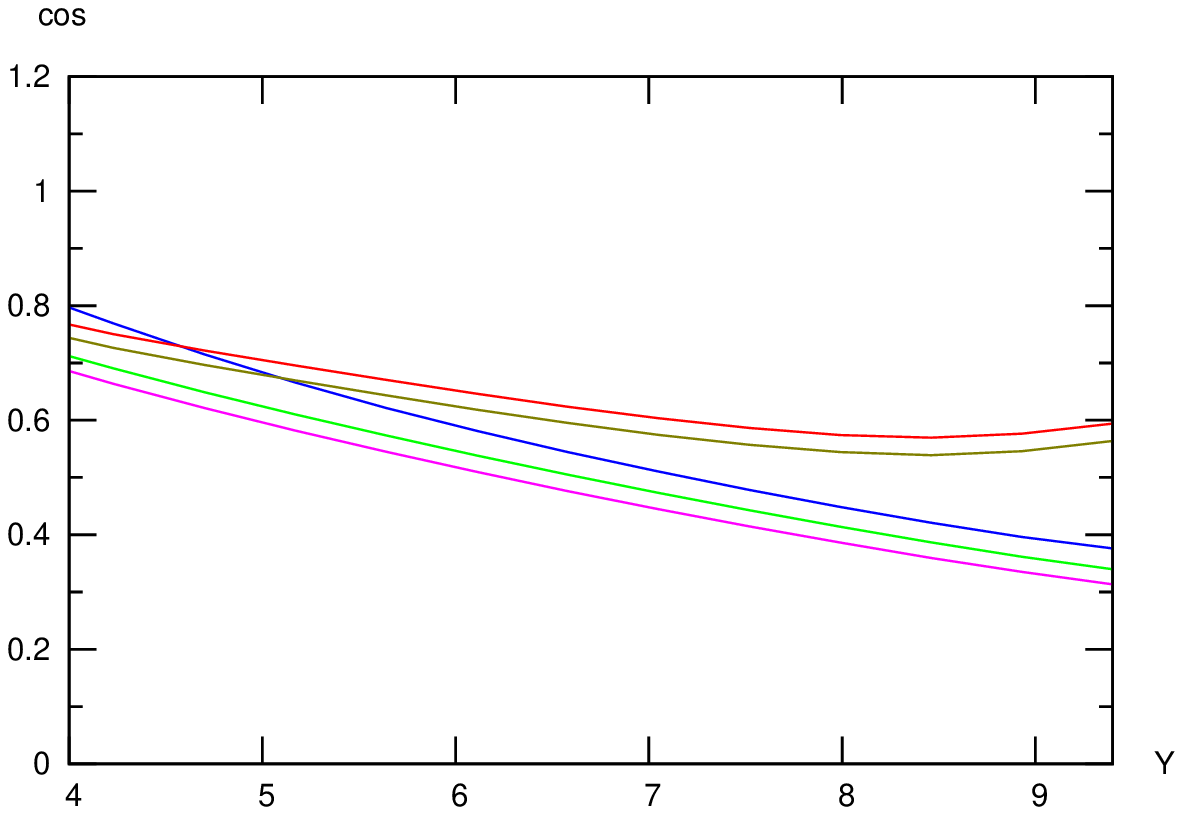}
  \end{minipage}
  \begin{minipage}{0.49\textwidth}
      \includegraphics[width=7cm]{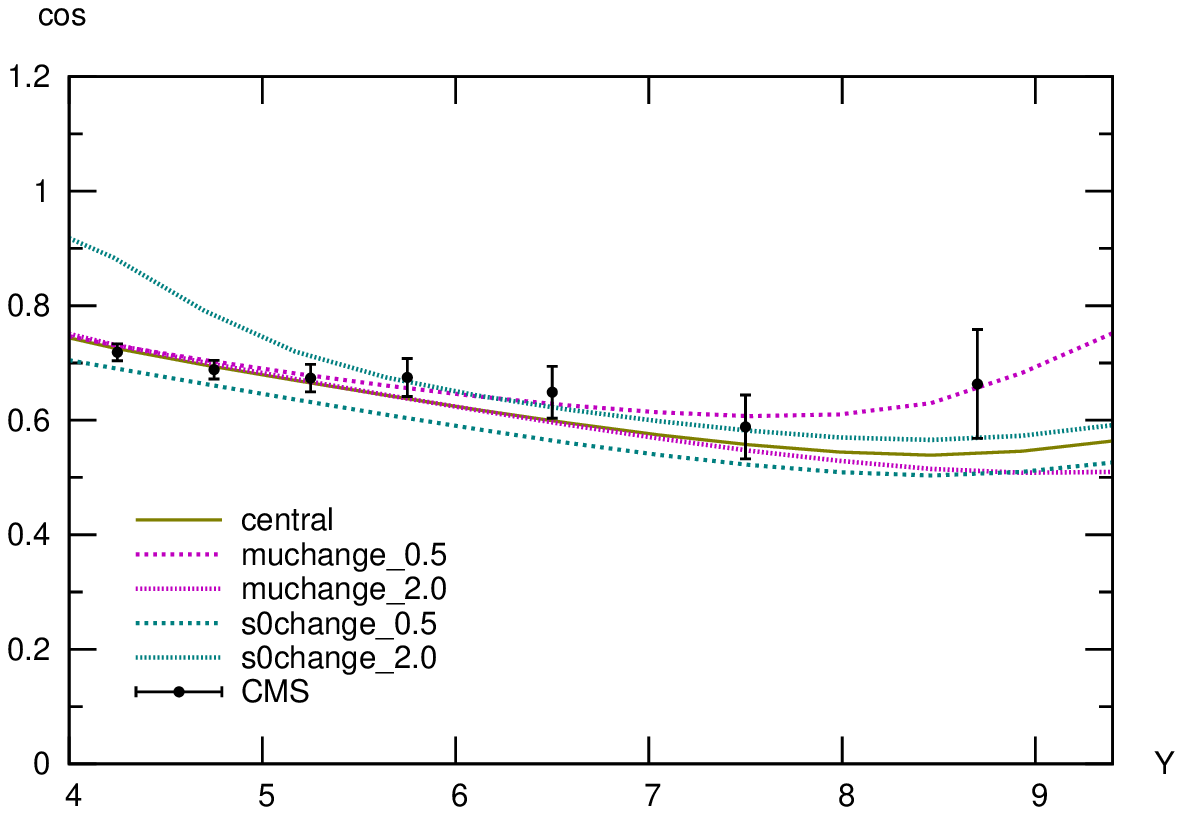}
  \end{minipage}
  \caption{Left: value of $\avgcostwo / \avgcos$ as a function of the rapidity separation $Y$, using symmetric cuts defined in (\protect\ref{sym-cuts}), for the 5 different BFKL treatments (\protect\ref{def:colors}). Right: comparison of the full NLL BFKL calculation including the scale uncertainty with CMS data (black dots with error bars).}
  \label{Fig:cos2cos_sym}
\end{figure}

The study of the ratio of the two observables studied above, $\avgcostwo/\avgcos$, was also done in~\cite{CMS-PAS-FSQ-12-002}. On figure~\ref{Fig:cos2cos_sym} we show our results for this observable. The impact of NLO corrections to the jet vertices is smaller than for $\avgcos$ and $\avgcostwo$ but still leads to a significantly different behavior than when using the LO vertices. On figure~\ref{Fig:cos2cos_sym} (R) we observe that the NLL BFKL calculation gives a good agreement with CMS data over the full $Y$ range and that this observable is more stable with respect to the scales than $\avgcos$ and $\avgcostwo$.

\section{Results for an asymmetric configuration}

The cuts chosen by the CMS collaboration in~\cite{CMS-PAS-FSQ-12-002} do not allow to perform a comparison with a fixed order calculation. Indeed, such calculations cannot give reliable results when the lower cut on the transverse momenta of the jets is the same for the two jets. Thus we cannot see if a BFKL calculation gives a better description of the data than a fixed order calculation, which would indicate that resummation effects have to be taken into account. Therefore, in this section we will compare our BFKL results with the results obtained using the NLO fixed order code \textsc{Dijet}~\cite{Aurenche:2008dn} with slightly different cuts, defined below:
\begin{eqnarray}
  35\,{\rm GeV} < &|\veckjone|, |\veckjtwo| & < 60 \,{\rm GeV} \,, \nonumber\\
  50\,{\rm GeV} < &{\rm Max}(|\veckjone|, |\veckjtwo|)\,, \nonumber\\
  0 < &y_{J,1}, \, y_{J,2}& < 4.7\,.
  \label{asym-cuts}
\end{eqnarray}

\begin{figure}[htbp]
  \def\sca{.6}
  \psfrag{central}[l][r][0.5]{\hspace{-0.9cm}pure NLL}
  \psfrag{muchange_0.5}[l][r][\sca]{\hspace{-1.9cm}\footnotesize $\mu_F \to \mu_F/2$}
  \psfrag{muchange_2.0}[l][r][\sca]{\hspace{-2cm} \footnotesize $\mu_F \to2 \mu_F$}
  \psfrag{s0change_0.5}[l][r][\sca]{\hspace{-2.05cm} \footnotesize $\sqrt{s_0} \to \sqrt{s_0}/2$}
  \psfrag{s0change_2.0}[l][r][\sca]{\hspace{-2.05cm} \footnotesize $\sqrt{s_0} \to 2 \sqrt{s_0}$}
  \psfrag{Dijet}[l][r][0.5]{\hspace{-0.7cm} fixed order NLO}
  \psfrag{cos}{\raisebox{.1cm}{\scalebox{0.9}{$\langle \cos \varphi\rangle$}}}
  \psfrag{Y}{\scalebox{0.9}{$Y$}}
  \begin{minipage}{0.49\textwidth}
    \hspace{-.3cm}  \includegraphics[width=7cm]{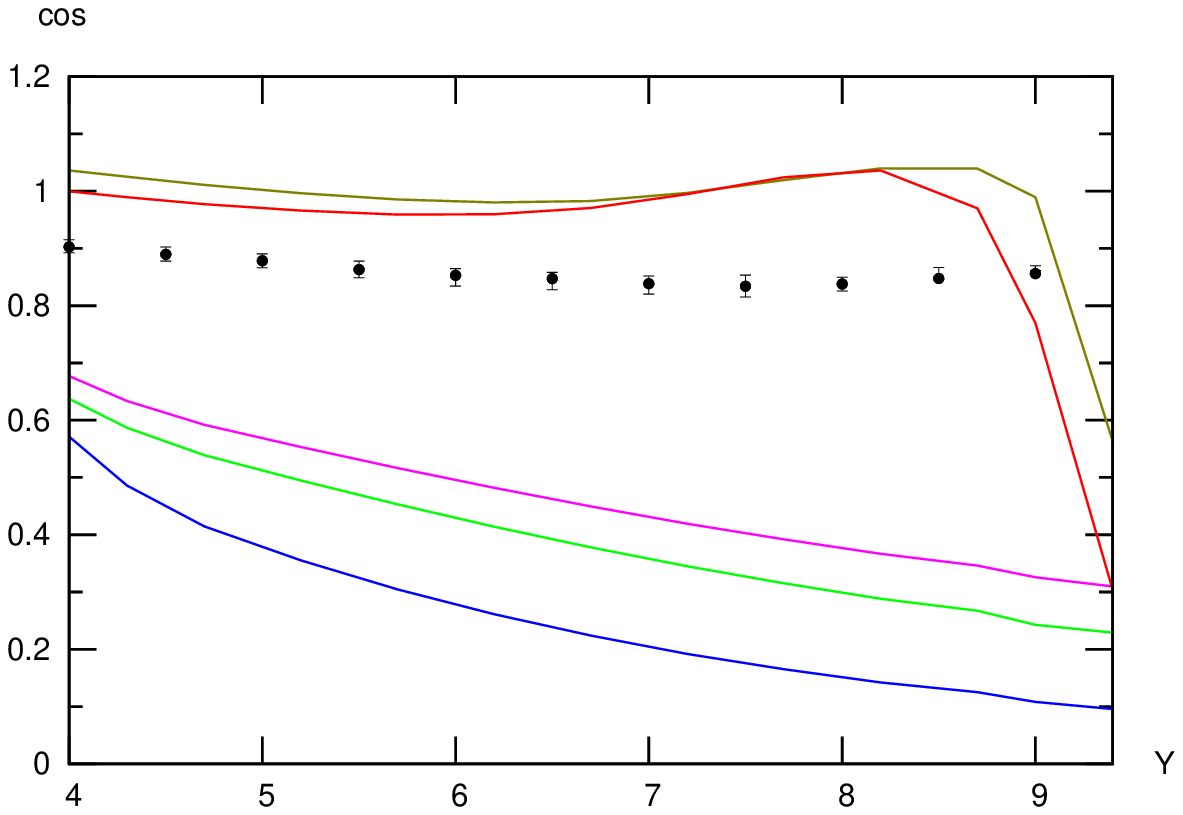}
  \end{minipage}
  \hspace{-.2cm}\begin{minipage}{0.49\textwidth}
      \includegraphics[width=7cm]{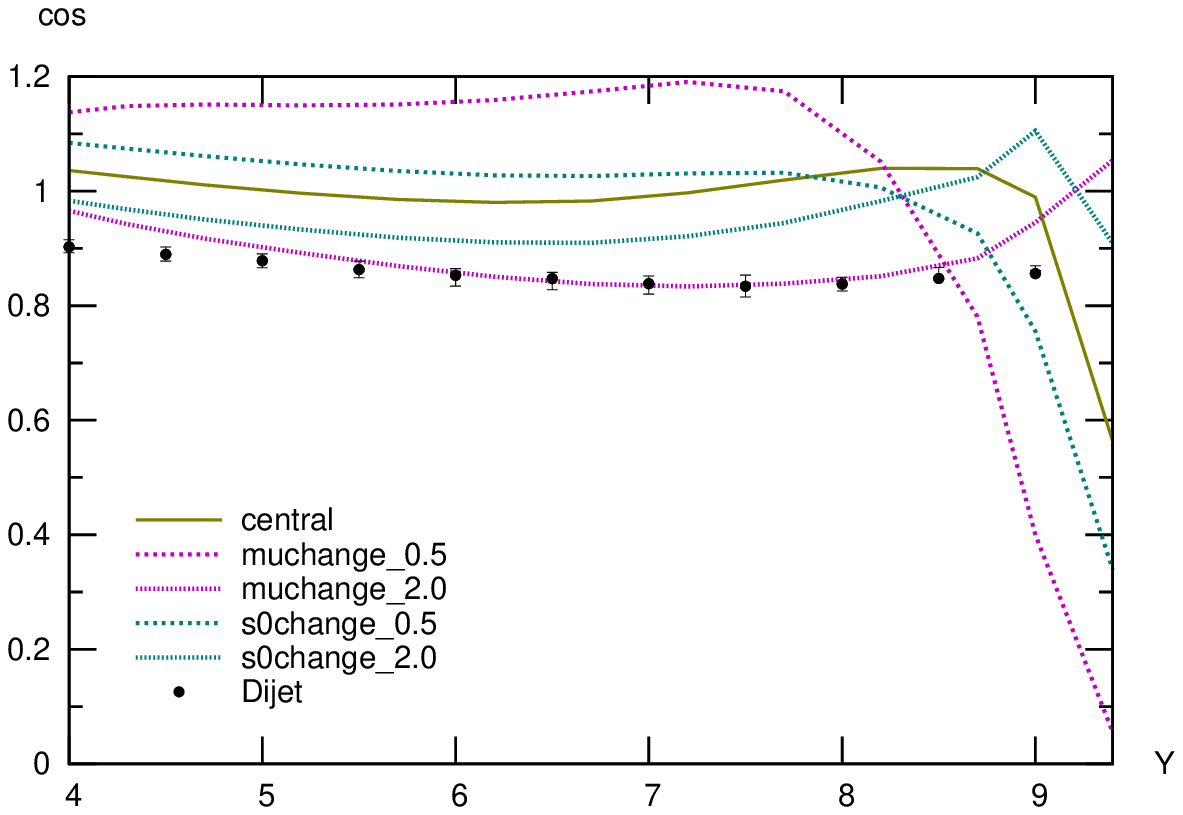}
  \end{minipage}
  \caption{Left: value of $\avgcos$ as a function of the rapidity separation $Y$, using asymmetric cuts defined in (\protect\ref{asym-cuts}), for the 5 different BFKL scenarios (\protect\ref{def:colors}). Right: comparison of the full NLL calculation including the scale uncertainty with \textsc{Dijet} predictions (black dots with error bars).}
  \label{Fig:cos_asym}
\end{figure}

The results of this comparison for the observable $\avgcos$ is shown on figure~\ref{Fig:cos_asym}. We see that the correlation predicted by \textsc{Dijet} (black dots with error bars, corresponding to varying the renormalization/factorization scale by a factor of $2$) is much larger than the three BFKL treatments using the LO vertices, and a little smaller than what we find in a full NLL calculation. But we can see that when varying the scales by a factor of 2, these predictions are compatible with each other.

\begin{figure}[htbp]
  \def\sca{.6}
  \psfrag{central}[l][r][0.5]{\hspace{-0.9cm}pure NLL}
  \psfrag{muchange_0.5}[l][r][\sca]{\hspace{-1.9cm}\footnotesize $\mu_F \to \mu_F/2$}
  \psfrag{muchange_2.0}[l][r][\sca]{\hspace{-2cm} \footnotesize $\mu_F \to2 \mu_F$}
  \psfrag{s0change_0.5}[l][r][\sca]{\hspace{-2.05cm} \footnotesize $\sqrt{s_0} \to \sqrt{s_0}/2$}
  \psfrag{s0change_2.0}[l][r][\sca]{\hspace{-2.05cm} \footnotesize $\sqrt{s_0} \to 2 \sqrt{s_0}$}
  \psfrag{Dijet}[l][r][0.5]{\hspace{-0.7cm} fixed order NLO}
  \psfrag{cos}{\raisebox{.1cm}{\scalebox{0.9}{$\langle \cos 2 \varphi\rangle$}}}
  \psfrag{Y}{\scalebox{0.9}{$Y$}}
  \begin{minipage}{0.49\textwidth}
    \hspace{-.3cm}  \includegraphics[width=7cm]{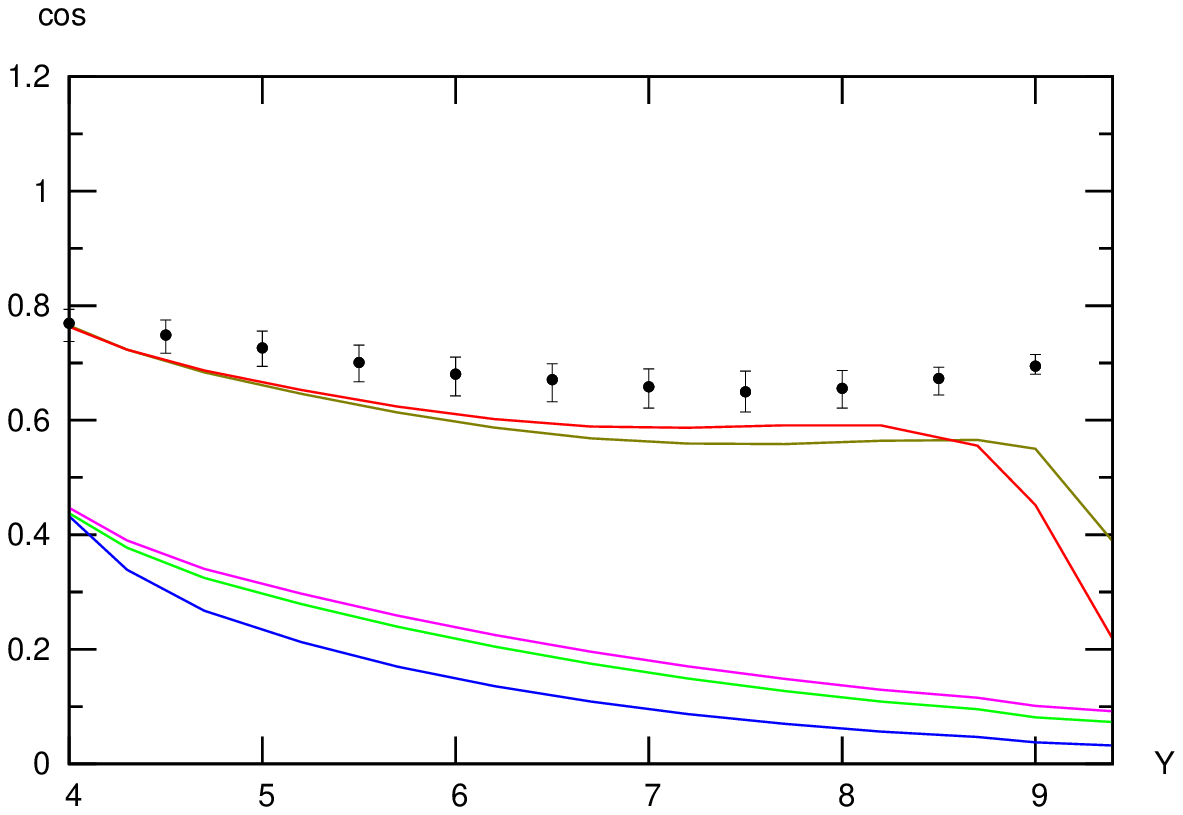}
  \end{minipage}
  \begin{minipage}{0.49\textwidth}
    \hspace{-.3cm}  \includegraphics[width=7cm]{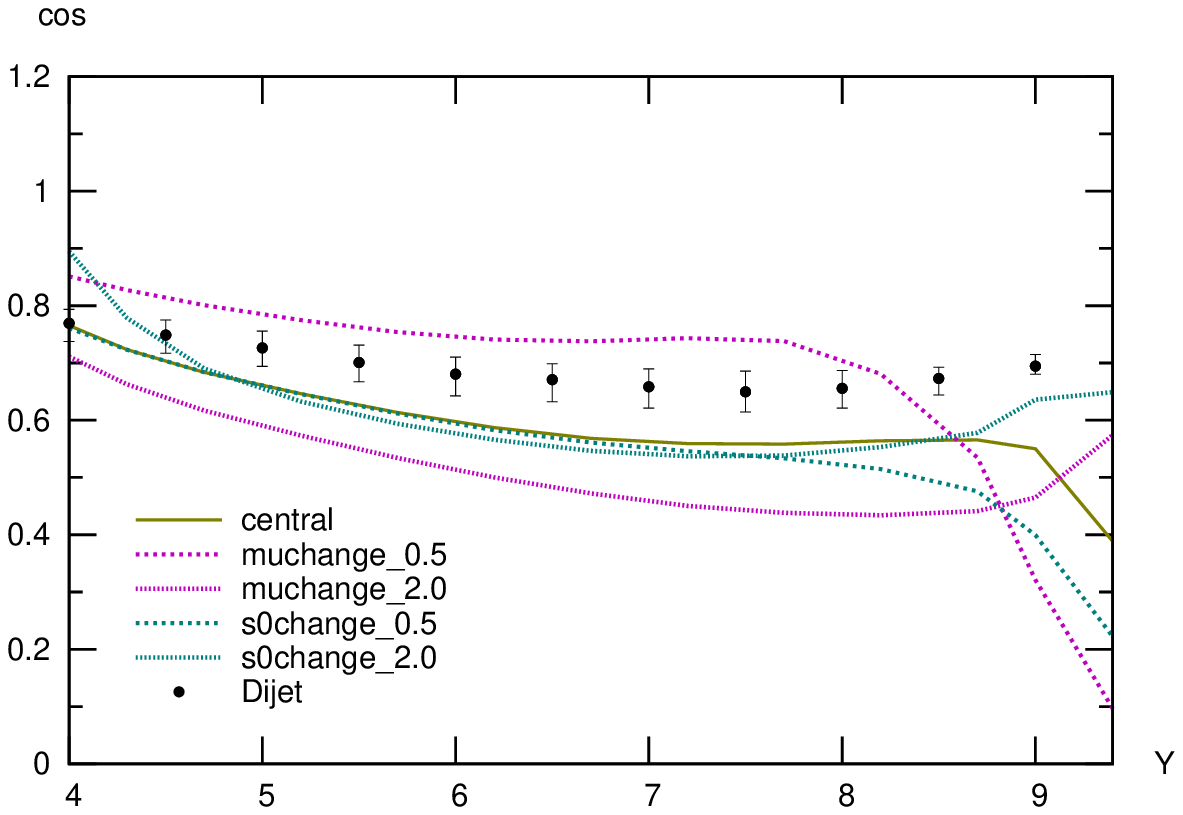}
  \end{minipage}
  \caption{Left: value of $\avgcostwo$ as a function of the rapidity separation $Y$, using asymmetric cuts defined in (\protect\ref{asym-cuts}), for the 5 different BFKL scenarios (\protect\ref{def:colors}). Right: comparison of the full NLL calculation including the scale uncertainty with \textsc{Dijet} predictions (black dots with error bars).}
   \label{Fig:cos2_asym}
\end{figure}

When we consider $\avgcostwo$, shown on figure~\ref{Fig:cos2_asym}, we observe that again the \textsc{Dijet} predictions are much closer to the BFKL treatments involving the NLO vertices than the ones involving the LO vertices. This time the values obtained with \textsc{Dijet} are slightly above our NLL results, but again we can see that these predictions are in compatible when taking into account the scale uncertainty.

\begin{figure}[htbp]
  \def\sca{.6}
  \psfrag{central}[l][r][0.5]{\hspace{-0.9cm}pure NLL}
  \psfrag{muchange_0.5}[l][r][\sca]{\hspace{-1.9cm}\footnotesize $\mu_F \to \mu_F/2$}
  \psfrag{muchange_2.0}[l][r][\sca]{\hspace{-2cm} \footnotesize $\mu_F \to2 \mu_F$}
  \psfrag{s0change_0.5}[l][r][\sca]{\hspace{-2.05cm} \footnotesize $\sqrt{s_0} \to \sqrt{s_0}/2$}
  \psfrag{s0change_2.0}[l][r][\sca]{\hspace{-2.05cm} \footnotesize $\sqrt{s_0} \to 2 \sqrt{s_0}$}
  \psfrag{Dijet}[l][r][0.5]{\hspace{-0.7cm} fixed order NLO}
  \psfrag{cos}{\raisebox{.1cm}{\scalebox{0.9}{$\langle \cos 2 \varphi\rangle / \langle \cos \varphi\rangle$}}}
  \psfrag{Y}{\scalebox{0.9}{$Y$}}
  \begin{minipage}{0.49\textwidth}
  \hspace{-.3cm}  \includegraphics[width=7cm]{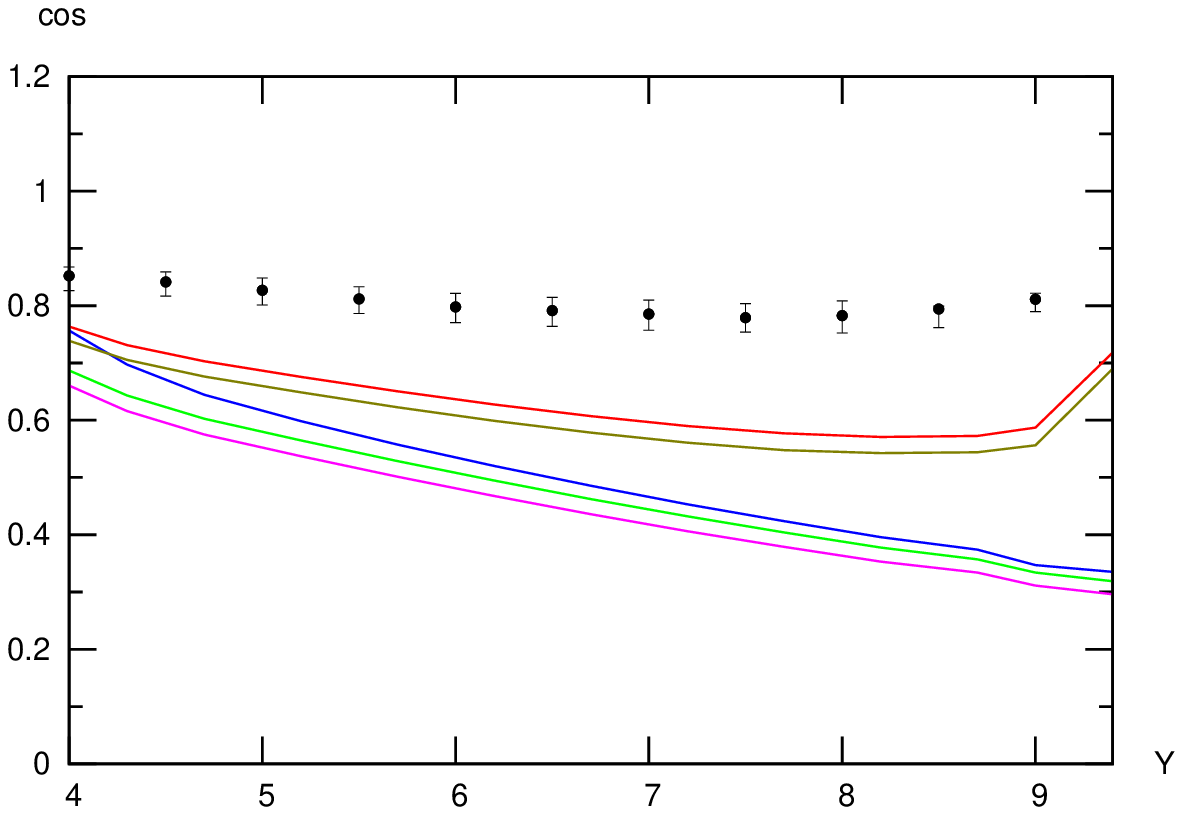}
  \end{minipage}
  \begin{minipage}{0.49\textwidth}
  \hspace{-.3cm}    \includegraphics[width=7cm]{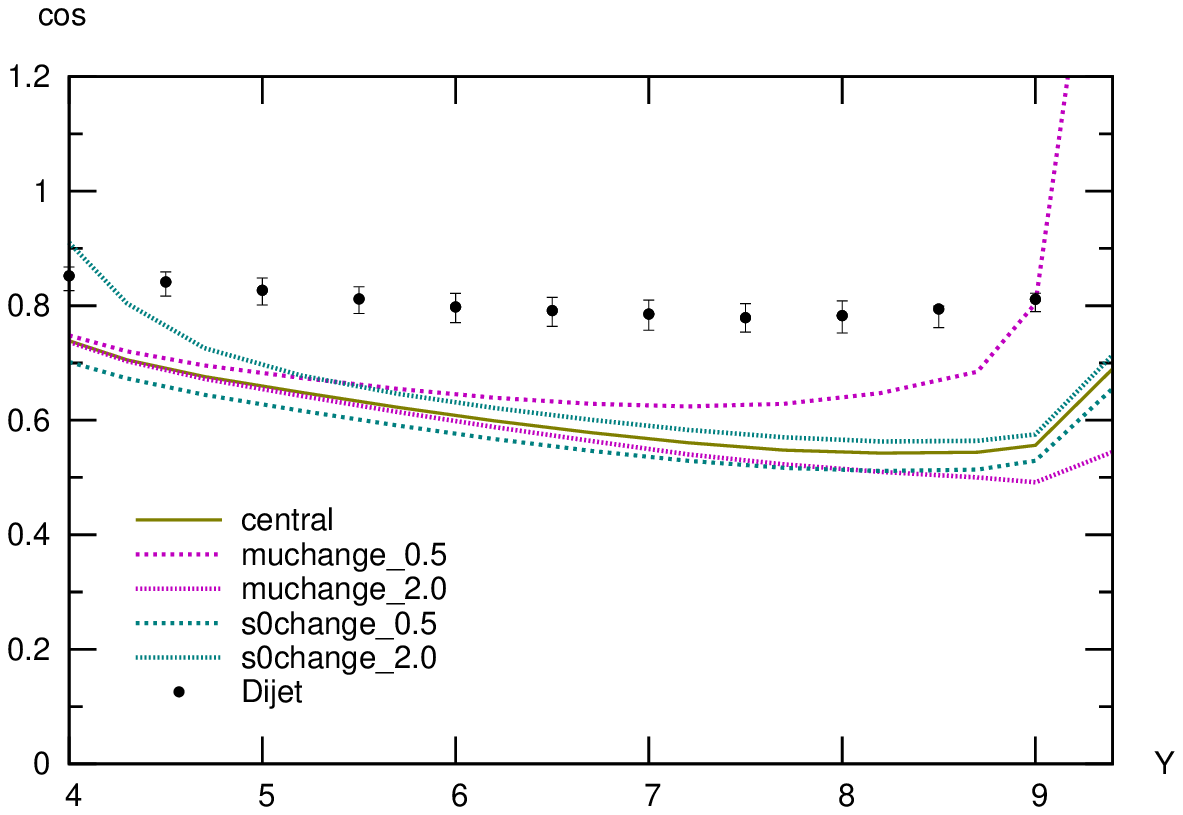}
  \end{minipage}
   \caption{Left: value of $\avgcostwo / \avgcos$ as a function of the rapidity separation $Y$, using asymmetric cuts defined in (\protect\ref{asym-cuts}), for the 5 different BFKL scenarios (\protect\ref{def:colors}). Right: comparison of the full NLL calculation including the scale uncertainty with \textsc{Dijet} predictions (black dots with error bars).}
\label{Fig:cos2cos_asym}
\end{figure}

We now do the same comparison for the observable $\avgcostwo / \avgcos$, shown on figure~\ref{Fig:cos2cos_asym}. We observe a significant difference between the BFKL treatments with LO and NLO vertices, and with NLO fixed order. As in the symmetric case, this observable is more stable with respect to the scales than $\avgcos$ or $\avgcostwo$ and the difference between NLL BFKL and NLO fixed order do not vanish when taking into account the scale uncertainty.

\section{Conclusions}

In this work we have compared the results of our full NLL BFKL calculation of azimuthal correlations of Mueller-Navelet jets to the first analysis performed at the LHC by the CMS collaboration. The comparison shows that for $\avgcos$ and $\avgcostwo$ a BFKL calculation using the LO jet vertices with the LL or NLL Green's function predicts a too large decorrelation and cannot describe the data. On the other hand, our results using the NLO jet vertices predict a too large correlation when compared to data, but the uncertainty associated with the choice of the scales is still quite large. 
We saw that the ratio of these observables is in good agreement with the data and more stable with respect to the scales.
Recently, we have shown~\cite{Ducloue:2013bva} that using the Brodsky-Lepage-Mackenzie~\cite{Brodsky:1982gc} procedure to fix the renormalization scale leads to a very good agreement of our calculation with experimental data for all the observables measured by the CMS collaboration.

It is not possible to confront the agreement of NLL BFKL and NLO fixed order calculations with the data, as the CMS collaboration chose the same lower cut for the transverse momenta of the jets, which is problematic for fixed order calculations. We thus compared our results with predictions of the NLO fixed order code \textsc{Dijet} in an asymmetric configuration. The outcome of this comparison is that the two approaches give compatible results for $\avgcos$ and $\avgcostwo$ when one takes into account the scale uncertainties. Contrary to these observables, $\avgcostwo/\avgcos$ is more stable with respect to the scales and significantly different results are found with NLL BFKL and NLO fixed order. Thus we believe that an experimental study in such an asymmetric configuration would be very useful to look for high energy resummation effects.

\acknowledgments

We thank Michel Fontannaz, Cyrille Marquet and Christophe Royon for
providing their codes and for stimulating discussions.
We warmly thank Grzegorz Brona, David d'Enterria, Hannes Jung, Victor Kim and
Maciej Misiura for many discussions and fruitful suggestions on the experimental aspects
of this study.

This work is supported by the French Grant PEPS-PTI, the
Polish Grant NCN No.~DEC-2011/01/B/ST2/03915 and  the Joint Research Activity Study of Strongly Interacting Matter (HadronPhysics3, Grant Agreement n.283286) under the 7th Framework Programm of the European Community.

\providecommand{\href}[2]{#2}\begingroup\raggedright\endgroup

%\bibliographystyle{JHEP}

%\bibliography{references-12-9-2013-2,publicationSW-source-10-9-2013,new_refs}

\end{document}